\documentclass[11pt]{article}
\usepackage{amssymb,epsf,graphicx}
\usepackage{amsmath}
\usepackage{epsfig}
\usepackage{latexsym}
\usepackage{graphicx}

\allowdisplaybreaks[1]

\begin{document}

\vskip 2.0cm

\centerline{\Large \bf Quantum gravity effects on statistics and
compact star configurations
}

\vspace*{8.0ex}

\centerline{\large Peng Wang\footnote{E-mail: {\tt
pengw@uestc.edu.cn}},   Haitang Yang\footnote{E-mail: {\tt
hyanga@uestc.edu.cn}} and Xiuming Zhang\footnote{E-mail: {\tt
zhangxm@uestc.edu.cn}}}

\vspace{2.5ex}

\centerline{\large \it Department of Applied Physics,}
\vspace{1.0ex}

\centerline{\large \it University of Electronic Science and
Technology of China,}

\vspace{1.0ex}

\centerline{\large \it Chengdu, 610054, People's Republic of China}

\vspace{3.0ex}

\vspace*{10.0ex}

\centerline{\bf Abstract}
\bigskip
The thermodynamics of classical and quantum ideal gases based on
the Generalized uncertainty principle (GUP) are investigated. At
low temperatures, we calculate corrections to the energy and
entropy. The equations of state receive small modifications. We
study a system comprised of a zero temperature ultra-relativistic
Fermi gas. It turns out that at low Fermi energy $\varepsilon_F$,
the degenerate pressure and energy are lifted. The Chandrasekhar
limit receives a small positive correction. We discuss the
applications on configurations of compact stars. As
$\varepsilon_F$ increases, the radius, total number of fermions
and mass first reach their nonvanishing minima and then diverge.
Beyond a critical Fermi energy, the radius of a compact star
becomes smaller than the Schwarzschild one. The stability of the
configurations is also addressed. We find that beyond another
critical value of the Fermi energy, the configurations are stable.
At large radius, the increment of the degenerate pressure is
accelerated at a rate proportional to the radius.

\vspace{3.0ex}


\vfill \eject

\baselineskip=16pt

\vspace*{10.0ex}

\tableofcontents

\section{Introduction}

It has been suggested that gravity itself leads to an effective
cutoff in the ultraviolet, i.e., a minimal observable length
\cite{Veneziano1986EPL199,DJGross1988NPB407,
DAmati1989PLB41,KKonishi1990PLB276,RGuida1991MPL1487,
Maggiore1993PLB304}. Some realizations of the minimal length from
various scenarios are proposed. One of the most important models
is the generalized uncertainty principle (GUP), derived from the
modified fundamental commutation relation
\cite{Maggiore1993PLB319,Garay1995IJMPA145, Kempf1995PRD1108,
Kempf1992LMP26,Kempf199234JMP969,
Kempf199235JMP4483,Hossenfelder2006CQG1815}
\begin{equation}
[x,p]=i\hbar(1+\beta p^2),\label{1dGUP}
\end{equation}
where $\beta=\beta_0\ell_p^2/\hbar^2=\beta_0/c^2M_p^2$ with the
Planck mass $M_p=\sqrt{\hbar c/G}$ and the Planck length
$\ell_p=\sqrt{{G\hbar}/{c^3}}$. $\beta_0$ is a dimensionless
parameter. With this generalization, one can easily derive the
generalized uncertainty principle (GUP)
\begin{equation}
\Delta x\Delta p\geq \frac \hbar 2 [1+\beta (\Delta
p)^2].\label{2dGUP}
\end{equation}
This in turn gives  the minimum measurable length
\begin{equation}
\Delta x\geq \Delta_{\textrm{min}}=\hbar\sqrt{\beta}
=\sqrt{\beta_0}\ell_p. \label{GUPshixian-2}
\end{equation}
Eqn. (\ref{1dGUP}) is the simplest model where only the minimal
uncertainty in position is taken into account while the momentum
can be infinite. In this case, the quantum-mechanical structure
underlying the GUP has been studied in full detail
\cite{Kempf1995PRD1108}. The planar waves in momentum space are
generalized by the maximal localization states and directly lead
to the modified de Broglie relation
\begin{equation}
\hbar k=\frac{\textrm{tan}^{-1}({\sqrt{\beta}}
p)}{{\sqrt\beta}}.\label{Mdr-1}
\end{equation}
In the limit $\beta\rightarrow 0$, one can obtain the usual de
Broglie relation. For $\beta\neq 0$, there is a nonzero minimal
wavelength
\begin{equation}
\lambda_0=4\hbar\sqrt\beta.
\end{equation}
Many implications and applications of  non-zero minimal length
have been discussed in literature \cite{Kempf1995PRD1108,
Hossenfelder2008CQG038003,RAMA2001PLB}. A more detailed list of
references refers to \cite{Fityo2008PLA}. In
\cite{DasVagenas2008PRL}, based on the precision measurement of
Lamb shift, an upper bound of $\beta_0$ is given by
$\beta_0<10^{36}$. A relatively rough but stronger restriction is
estimated in \cite{FBrauB2006PRD036002}.
Furthermore, in \cite{MuWuYang2009print}, the parameter $\beta_0$
is conjectured to vary with energy scales. In our analysis,
$\beta_0<10^{36}$ is adopted.

In this paper, we discuss the thermodynamic properties of ideal
gases based on GUP. The statistical physics of ideal gases has
been studied by many authors
\cite{RAMA2001PLB,Fityo2008PLA,LNC2002PRD,
RAMA2001-0204215,NozariMehdipour2007Chaos}. It turns out that the
thermodynamic quantities at finite temperatures are shifted by
$\beta k_\textrm{B}T$ for non-relativistic system or $\beta
(k_\textrm{B}T)^2$ for ultra-relativistic and photon ones.
Accordingly, there exist modifications on the equations of state.

It is well known that the cold relativistic Fermi gas has
important applications in astrophysics. Especially in 1930s,
Chandrasekhar found that white dwarfs are very well described in
the framework of a highly degenerate ideal electron gas.
Therefore, it is of interest to investigate the ultra-relativistic
ideal Fermi gas at zero temperature. It proves that both the
energy and pressure of the system receive quantum gravity
corrections as the Fermi energy is low. Since a full theory of
quantum gravity is absent, it is a good try to take GUP model as a
starting point and apply it to the case of high Fermi energy. We
find that both the number density and the energy density approach
finite values, whereas the degenerate pressure blows up.

We are going to address two kinds of compact star configurations.
One is white dwarf alike where the major contribution of the mass
is not from the Fermi gas under discussion but from the cold
nuclei. Application of our arguments to this type of compact stars
causes small corrections to the Chandrasekhar limit. The mass of
another configuration is expressed as $M= U_0/c^2$, where the
compact star is completely constituted by an ideal Fermi gas. Our
calculation shows that the radius, total number and mass all have
minima at Fermi energies around $E_{\rm H} = M_p
c^2/\sqrt\beta_0$. The minimum radius is proportional to
$\sqrt\beta_0 \Delta_{\rm min}$, different from the intuitively
expected value $\Delta_{\rm min}$. The radius becomes smaller than
the Schwarzschild one beyond a critical value of the Fermi energy,
though eventually approaches infinity. This scenario provides a
possible interpretation of our universe as a giant black hole. We
further find that the degenerate pressure increases at a rate
proportional to the radius asymptotically. A question therefore
naturally arises: Is the accelerating expansion of our universe
accounted by Fermi pressure?

We consider an isolated macroscopic body, consisting of $N$
non-interacting particles. In condensed matter physics, the
background of particles' motion is flat. However, as we know from
(\ref{1dGUP}) and (\ref{GUPshixian-2}), the generalized
uncertainty principle realizes that the particles move in
quantized gravitational background. It is then natural to consider
the nearly independent particle systems. The state density has
been derived in several ways
\cite{RAMA2001PLB,Fityo2008PLA,LNC2002PRD}. In \cite{LNC2002PRD},
with the help of Liouville theorem, the density of states is given
\begin{equation}
\frac{Vdp_1dp_2\cdots dp_d}{h^d(1+\beta
p^2)^d}.\label{statenonumber-LNC1}
\end{equation}
In $d$-dimensional spherical coordinate systems, the state density
in momentum space is
\begin{equation}
D(p)dp=\frac{VA(S^{d-1})p^{d-1}dp}{h^d(1+\beta p^2)^d},
\;\;p\in(0,+\infty),\label{statenonumber-qiuzuobLNC1}
\end{equation}
where $A(S^{d-1})={2\pi^{d/2}}/{\Gamma(\frac{d}2)}$, with
$\Gamma(\frac d2)$ being the Gamma function. If the particle's
spin is 1 or $1/2$, the two equations above should be multiplied
by $2$.

Let's review the organization of this paper. In section 2 we
discuss the finite temperature classical non-relativistic,
ultra-relativistic and photon gases respectively. In section 3, we
explore the properties of an ultra-relativistic Fermi gas at zero
temperature. Section 4 is devoted to the discussion of  compact
star configurations. In section 5, we offer a summary and
discussion.

\section{Ideal gases at finite temperatures}
The ideal gas model plays a major role in the studies of
statistics. When the electromagnetic interactions between
particles can be neglected, the ideal gas is a good approximation
for a real system. In this section, we discuss low temperature
classical non-relativistic, ultra-relativistic systems and a
photon gas respectively.

\subsection{classical statistics}

For a non-relativistic system, the dispersion relation is
$\varepsilon=p^2/2m$. From (\ref{statenonumber-qiuzuobLNC1}), in
$d$-dimensional space, the partition function is
\begin{eqnarray}
Z&=&\frac{VA(S^{d-1})}{h^d}\int_0^{\infty}\frac{p^{d-1}
e^{-\frac{\beta_{\textrm{B}}}{2m}p^2}dp}{(1+\beta
p^2)^d}\nonumber
\\
& =&\frac{VA(S^{d-1})}{h^d}(2mk_\textrm{B}T)^{\frac
d2}I(\gamma,d), \label{ddpaetitionfN-11}
\end{eqnarray}
where $\gamma\equiv 2m\beta/\beta_\textrm{B}=2m\beta
k_\textrm{B}T$ is a dimensionless parameter in the natural unit
system and $\beta_B = 1/k_B T$ with $k_\textrm{B}$ being the
Boltzmann's constant. $I(\gamma,d)$ is defined as
\begin{eqnarray}
I(\gamma,d)&&=\frac{2^{-d}\sqrt{\pi}\; \gamma^{-d/2 }\;
   \Gamma[\frac d2] H\left[\frac d2, 1 - \frac d2, \frac 1\gamma\right]}
   {\Gamma[\frac {d+1}2] }
+\frac 12 \gamma^{-d} \Gamma[-\frac d2] H\left[d, 1 + \frac d2,
\frac 1\gamma\right]\nonumber\\
&&\approx \frac 12\Gamma\left[  \frac{d}{2}\right] \left(
1-\frac{d^{2}}{2}\gamma\right)  +O\left(  \gamma^{2}\right),
\label{IGAMMAD}
\end{eqnarray}
where $H[\;,\;,\;]$ is the Kummer confluent hypergeometric
function. It is straightforward to derive the equation of state of
the ideal gas $PV=Nk_\textrm{B}T $ from eqn.
(\ref{ddpaetitionfN-11}). For simplicity, we consider $d=3$,
\begin{equation}
I(\gamma\ll 1,3)=\frac{\sqrt{\pi}}{4}\left[1-\frac
92\gamma\right]+O(\gamma^2).\label{IGAMAMAQD3XA-1}
\end{equation}

In \cite{DasVagenas2008PRL}, an upper limit of $\beta_0$ is
estimated as $\beta_0<10^{36}$ or $\beta<10^{34}(\frac{{\rm s}}{{\rm
kg}\cdot {\rm m}})^2$. If choosing $m$ to be the electron mass,
$m=10^{-30}\textrm{kg}$, one has
$$
\frac{\gamma}{T}<{10^{-19}}{\textrm{K}^{-1}}.
$$
In the case of low temperature, $0<T\ll
10^{19}\textrm{K}$, i.e., $\gamma \ll 1$, from eqn.
(\ref{ddpaetitionfN-11}) and eqn. (\ref{IGAMAMAQD3XA-1}), we have
\begin{equation}
Z\approx\frac{4\pi V}{h^3}(2mk_\textrm{B}T)^{3/2}\frac{\sqrt
\pi}{4}\left[1-\frac 92\gamma\right].
\end{equation}
When $\beta=0$, one recovers the canonical energy $U_0=\frac
32Nk_\textrm{B}T$ and the entropy
\begin{equation}
S_0=Nk_\textrm{B}\left[\frac{5}2+\frac 32\,\textrm{ln}\frac{2\pi
mk_\textrm{B}T}{h^2}+\textrm{ln}\frac VN\right].
\end{equation}
We are concerned with  $\beta\neq 0$ where the quantum gravity
effects are introduced. Combined with $0<T\ll 10^{19}\textrm{K}$,
the energy gains a small correction
\begin{equation}
U=\frac 32Nk_\textrm{B}T+\delta U,\;\; \delta U=-\frac 92
Nk_\textrm{B}T\,2\beta mk_\textrm{B}T.\label{ddenergynz-1}
\end{equation}
Notice that the energy has nothing to do with the volume. This is
entirely consistent with the case $\beta=0$. The entropy is also
modified,
\begin{eqnarray}
S&=& S_0+\delta S,\nonumber\\
\delta S&=&\frac{\delta U}{T}+Nk_\textrm{B}\textrm{ln}\left[1-\frac
92\gamma\right]=-18Nk_\textrm{B}^2mT\beta. \label{NonreS-1}
\end{eqnarray}

For a nonrelativistic system, one can calculate the Maxwellian
distribution in $d>1$ at low temperatures. Using
(\ref{statenonumber-qiuzuobLNC1}), the velocity probability
distribution for the absolute magnitude in $d$-dimensional space
takes the form
\begin{equation}
d\omega_{\emph{{v}}}\sim\frac{e^{-{mv^{2}}/{2k_\textrm{B}T}}}
{{(1+\beta m^2v^{2})^d}}v^{d-1}dv.\nonumber
\end{equation}
The most probable distribution of the velocity is determined by
the extremum,
\begin{eqnarray*}
v_\textrm{m}^2=\frac
{\sqrt{\left[(d+1)\gamma+2\right]^2+8(d-1)\gamma}-
\left[(d+1)\gamma+2\right]}{4\gamma}\approx \frac
{(d-1)k_\textrm{B}T}{m}-2d\beta mk_\textrm{B}T.
\end{eqnarray*}
Comparing with the classical statistics, there is  a negative
shift proportional to $\beta k_\textrm{B}T$.

When the particle's rest mass is very small or the temperature is
high enough, the condition $k_\textrm{B}T\gg mc^2$ is well
satisfied. It is natural to study the ultra-relativistic ideal
gas with the dispersion $\varepsilon=cp$. 
Then from
(\ref{statenonumber-qiuzuobLNC1}), we have
\begin{eqnarray}
Z&=&\frac{VA(S^{d-1})}{h^d}\int_0^{\infty}\frac{p^{d-1}
e^{-{\beta_{\textrm{B}}}cp}dp}{(1+\beta
p^2)^d}\nonumber\\
& =&\frac{VA(S^{d-1})}{h^d}\left(\frac{k_\textrm{B}T}{c}\right)^{d}
\int_0^{\infty}\frac{x^{d-1}e^{-x}dx}{(1+qx^2)^d}\nonumber\\
&\equiv& \frac{VA(S^{d-1})}{h^d}\left(\frac{k_\textrm{B}T}
{c}\right)^{d}\tilde{I}(q,d), \label{jiduanxiangduilpeifhans1d-1}
\end{eqnarray}
where $q\equiv\beta(k_\textrm{B}T)^2/c^2$ is a dimensionless
parameter. For generic dimensions, $\tilde{I}(q,d)$ is expressed
as complicated generalized hypergeometric functions. Therefore, we
still focus our attention on $d=3$,
\begin{equation}
\tilde{I}(q,3)=\frac{M\left[\{\{-\frac 12\},\{\}\},\{\{0,\frac
12,\frac 32\},\{\}\},\frac 1{4q}\right]}{4\sqrt\pi\,q^{3/2}},
\end{equation}
where $M[\;,\;] $ is  the Meijer G function. For  $q\ll 1$, we
find
\begin{equation}
\tilde{I}(q\ll 1,3)=2-72q+O(q^2).\label{jiduanxijifhansh1-1}
\end{equation}
One can see that $q= 1$ defines a temperature $k_{\rm B}T_q =
\frac{c}{\sqrt\beta}$. It is believed that the minimum  length is
about the length of a string: $\Delta_{\rm min} = \hbar
\sqrt\beta\simeq \ell_s$. Therefore,
\[k_{\rm B}T_q  \simeq \frac{c\hbar}{\ell_s} = 4\pi k_{\rm B}T_{\rm H},\]
where $T_{\rm H}$ is the  Hagedorn temperature of relativistic
strings
\cite{AtickWitten1988NPB310,BZwiebach2004firstcoursestring16}. On
the other hand, the Hagedorn temperature is estimated to be
$T_{\rm H}\simeq 10^{30}$K by the traditional grand-unified string
models. Thus, a lower bound of $\beta_0$ is imposed by $T_{\rm
H}$:
\[\beta_0> 10^4.\]
From now on, we name $k_{\rm B}T_q = \frac{c}{\sqrt\beta} = M_p
c^2/\sqrt\beta_0$ as Hagedorn energy or Hagedorn temperature, the
scale where the quantum gravity effects become important. When the
temperature is far below the Hagedorn temperature  $q \ll 1$, for
$d=3$, similar to
(\ref{ddenergynz-1}) and (\ref{NonreS-1}), 
the energy and entropy acquire small corrections due to the quantum gravity effects,
\begin{equation}
U=3Nk_\textrm{B}T\left[1-24\frac\beta{c^2}(k_\textrm{B}T)^2\right],
\label{ExRelCU}
\end{equation}
\begin{equation}
S=3Nk_\textrm{B}+Nk_\textrm{B}\textrm{ln}\left[\frac{8\pi
V}{h^3}\left(\frac{k_\textrm{B}T}{c}\right)^3\right]-k_\textrm{B}\textrm{ln}N!
-108Nk_\textrm{B}\frac{\beta}{c^2}(k_\textrm{B}T)^2.
\label{ExRelCS}
\end{equation}

In (\ref{ddpaetitionfN-11}) and
(\ref{jiduanxiangduilpeifhans1d-1}), the integrands contain
exponential functions. Given $\gamma\ll 1$ or $q\ll 1$, the
integrals are dominated by the neighborhoods of the maxima of the
numerators. These maxima are the order of unity. Therefore, one
can expand the denominators of the integrands before performing
integrations. Employing this method, the integral
(\ref{jiduanxiangduilpeifhans1d-1}) takes the form
\begin{equation}
Z\approx
\frac{VA(S^{d-1})}{h^d}\left(\frac{k_\textrm{B}T}{c}\right)^{d}
[\Gamma(d)-d\,\Gamma(d+2)q],
\end{equation}
which precisely agrees with eqn. (\ref{jiduanxijifhansh1-1}) for $d=3$. The
non-relativistic one (\ref{IGAMMAD}) is also reproduced by this method.
We are going to apply this method in the
following discussion of the photon gas.

\subsection{The photon gas}

In $d$-dimensional space, given $q=\beta (k_B T)^2/c^2 \ll 1$, the
grand partition function of the photon gas is
\begin{eqnarray}
\textrm{ln}\Xi&=&-2\int_0^\infty\frac{
VA(S^{d-1})\varepsilon^{d-1}}{(hc)^d(1+
\frac{\beta}{c^2}\varepsilon^2)^d}\textrm{ln}(1-e^{-\beta_\textrm{B}
\varepsilon})d\varepsilon\nonumber\\
&\approx&2\frac{ VA(S^{d-1})}{(hc)^d
\beta_\textrm{B}^{d}}\left[\frac{1}{d}\;\zeta(d+1)\Gamma(d+1)
-\frac{q\, d}{(d+2)}\;\zeta(d+3)\Gamma(d+3)\right],
\label{partitionfphoton-1}
\end{eqnarray}
where $\zeta (d)$ is the Riemann Zeta function. The energy is
\begin{equation}
U
=\frac{
2VA(S^{d-1})}{(hc)^d}(k_\textrm{B}T)^{d+1}\;\zeta(d+1)\Gamma(d+1)\cdot
\left[1-d\, q\,\frac{\zeta(d+3)
\Gamma(d+3)}{\zeta(d+1)\Gamma(d+1)}\right]\nonumber\\
\equiv U_0+\delta U.\label{guangziqiten-1}
\end{equation}
The entropy reads
\begin{equation}
S=k_\textrm{B}\textrm{ln}\Xi+\frac UT=S_0+\delta S,
\end{equation}
with
\begin{eqnarray}
\delta S =-q\,k_\textrm{B}\,\frac{
2VA(S^{d-1})}{(hc)^d}\;\frac{d(d+3)}{d+2}\;\zeta(d+3)\Gamma(d+3)
(k_\textrm{B}T)^{d}.\label{guangzishang-1}
\end{eqnarray}
From (\ref{ddenergynz-1}), (\ref{NonreS-1}), (\ref{ExRelCU}),
(\ref{ExRelCS}), (\ref{guangziqiten-1}) and
(\ref{guangzishang-1}), one can check that
\begin{equation}
\frac{\partial S}{\partial U}=\frac{\partial S_0}{\partial
U_0}+\frac{\partial\delta S}{\partial U_0}-\frac{\partial\delta
U}{\partial U_0}\frac{\partial S_0}{\partial U_0}=\frac{\partial
S_0}{\partial U_0}.\label{wenduxiuzheng1}
\end{equation}
In thermodynamics, there are only three independent basic
thermodynamic quantities, the temperature $T$, inner energy $U$
and entropy $S$. After introducing  GUP, $S$ and $U$ are altered,
while eqn. (\ref{wenduxiuzheng1})
shows that the temperature does not change. 

\section{Ground state properties of ultra-relativistic Fermi gases}

From the previous discussion, one can see that the deformed
parameter $\beta$ and the temperature $T$ are independent. What
really matters is $\beta T$ or $\beta T^2$ respectively. In this
section, we study the ground state properties of a Fermi gas
composed of $N$ ultra-relativistic electrons. For non-interacting
and ultra-relativistic particles, $\varepsilon=cp$. For simplicity
we only consider the space dimension $d=3$. When the temperature
is higher than the particle's rest mass (in the natural units),
particle pairs could be  produced. In this case the grand
canonical distribution oughts to be used. However, at zero or low
temperatures, the vacuum effect of fermions can be neglected.
Therefore, the total particle number is conserved. At zero
temperature, from (\ref{statenonumber-qiuzuobLNC1}), the Fermi
energy $\varepsilon_F$  is given by
\begin{eqnarray}
N&=&\frac{8\pi V}{h^3}\int_0^{p_F} \frac{p^{2}dp}{(1+{\beta}
p^2)^3}=\frac{8\pi V}{(hc)^3}\int_0^{\varepsilon_F}
\frac{\varepsilon^{2}d\varepsilon}{(1+\frac{\beta}{c^2}
\varepsilon^2)^3}\nonumber\\
&=&\frac{8\pi V}{(hc)^3}\left(\frac{c^2}{\beta}\right)^{3/2}\frac
1{8}\left[\frac{\kappa(\kappa^2-1)}{(1+\kappa^2)^2}+
\textrm{tan}^{-1}(\kappa)\right]\nonumber\\
 &\equiv&\frac{8\pi
V}{(hc)^3}E_{\rm H}^3\,f(\kappa), \label{fermilizishu-1}
\end{eqnarray}
where we defined $\kappa=\varepsilon_F
\sqrt{\frac{\beta}{c^2}}=\varepsilon_F  /E_{\rm H} $ with $E_{\rm
H} $ being the Hagedorn energy introduced in last section and
\begin{equation}
f(\kappa)=\frac
1{8}\left[\frac{\kappa(\kappa^2-1)}{(1+\kappa^2)^2}+
\textrm{tan}^{-1}(\kappa)\right]. \label{fDefinition}
\end{equation}
Then the average distance between particles is
\begin{equation}
\bar{d}\equiv\left(\frac{V}{N}\right)^{1/3}=\frac{hc}{(8\pi)^{1/3}}
\left(\frac{\beta}{c^2}\right)^{1/2} f(\kappa)^{-\frac 13}
=\pi^{\frac 23}\Delta_{\textrm{min}}f(\kappa)^{-\frac
13}.\label{adp-1feirm}
\end{equation}
The ground-state energy of the system is
\begin{eqnarray}
U_0&=&\frac{8\pi V}{(hc)^3}\int_0^{\varepsilon_F}
\frac{\varepsilon^{3}d\varepsilon}{(1+\frac{\beta}{c^2}
\varepsilon^2)^3}\nonumber\\
&=&\frac{8\pi V}{(hc)^3}\;\frac
14\varepsilon_F^4\frac{1}{(1+\kappa^2)^2}\label{fermijitaineng-123}\\
&=&\frac{8\pi V}{(hc)^3}\;E_{\rm H}^4 \frac
14\frac{\kappa^4}{(1+\kappa^2)^2}\label{fermijitaineng-124}.
\end{eqnarray}
The pressure of the system is given by
\begin{equation}
P=\frac{8\pi}{(hc)^3}k_\textrm{B}T\int_0^\infty
\frac{\varepsilon^2d\varepsilon}{(1+\frac{\beta}{c^2}\varepsilon^2)^3}
\;\textrm{ln}\left(1+e^{-\frac{(\varepsilon-\mu)}{k_\textrm{B}T}}
\right).\label{yibanyaqfeiji-1234}
\end{equation}
At zero temperature, the pressure becomes
\begin{equation}
P_0=\frac NV \varepsilon_F-\frac
{U_0}V.\label{yaqheneinengguanxi-1123ee21}
\end{equation}
Plugging in (\ref{fermilizishu-1}) and (\ref{fermijitaineng-124}),
one has
\begin{equation}
P_0= \frac{8\pi}{(hc)^3} E_{\rm H}^4 \,g(\kappa), \quad \quad
g(\kappa) \equiv\kappa f(\kappa)-\frac
14\frac{\kappa^4}{(1+\kappa^2)^2}.\label{yaqheneinengguanxi-1121}
\end{equation}
It is easy to see that $P_0$ is a monotonically increasing
function of $\kappa$ in the region $\kappa>0$.

The equation of state plays an indispensable role in exploring the
configuration of stars and cosmology.
Unfortunately, an explicit expression of the pressure in terms of
$\beta$ and $N/V$ is not reachable with eqn.
(\ref{fermilizishu-1}) and eqn. (\ref{yaqheneinengguanxi-1121}).
Therefore, we address the extremal situations in the following
discussions.

\subsection{$\kappa\ll 1$ case}

In this situation, 
\begin{equation}
f(\kappa)= \frac{\kappa^3}{3} -\frac{3}{5} \kappa^5 +
O(\kappa^7).
\label{FConvertKappa}
\end{equation}
This equation enables us to solve $\kappa$ in terms of $N/V$ and
$\beta$ from eqn. (\ref{fermilizishu-1}):
\begin{equation}
\kappa = (3\pi^2)^{1/3}\delta \left(1+ \frac{3}{5} (3\pi^2)^{2/3}
\delta^2\right)+ O(\delta^5),
\end{equation}
where we have set $\delta\equiv{\Delta_\textrm{min}}/{\bar{d}}=
(N/V)^{1/3} \hbar\sqrt\beta$. It is easy to see that $\delta \ll
1$ in this case from (\ref{adp-1feirm}). The Fermi energy is far
below the Hagedorn energy,
\begin{equation}
\varepsilon_F\ll E_{\rm H}=
{\frac{c^2}{\sqrt{\beta_0}}}\,M_p\;.\label{feimijinsitia-1}
\end{equation}
With equations
(\ref{fermijitaineng-124}) and
(\ref{yaqheneinengguanxi-1121}), one  obtains
\begin{eqnarray}
U_0/V&=& \frac{8\pi }{(hc)^3}\left(\frac{N}{8\pi V}\right)^{\frac
43}(hc)^4\cdot\frac 14\frac
{\kappa^4}{(1+\kappa^2)^2}f(\kappa)^{-\frac
43}\nonumber\\
&\approx& \frac{3^{4/3}}{4} \frac{hc}{(8\pi)^{1/3}}
\left(\frac{N}{V}\right)^{4/3} \left[1 + \frac{2}{5} \kappa^2
\right] .
\end{eqnarray}
\begin{eqnarray}
P_0&=&\frac{8\pi }{(hc)^3}\left(\frac{N}{8\pi V}\right)^{\frac
43}(hc)^4\cdot f(\kappa)^{-\frac 43}\left[\kappa
 f(\kappa)-\frac 14\frac{\kappa^4}{(1+\kappa^2)^2}\right].\nonumber\\
&\approx& \frac{3^{1/3}}{4} \frac{hc}{(8\pi)^{1/3}}
\left(\frac{N}{V}\right)^{4/3} \left[1 + \frac{6}{5} \kappa^2
\right] .
\end{eqnarray}
Replacing $\kappa$ by $\delta$, the energy and pressure both
receive corrections:
\begin{eqnarray}
&&U_0/V\approx \frac{3^{4/3}}{4} \frac{hc}{(8\pi)^{1/3}}
\left(\frac{N}{V}\right)^{4/3} \left[1 + \frac{2}{5}
(3\pi^2)^{2/3} \delta^2
\right],\\
 && P_0\approx\frac{3^{1/3}}{4} \frac{hc}{(8\pi)^{1/3}}
\left(\frac{N}{V}\right)^{4/3} \left[1 + \frac{6}{5}
(3\pi^2)^{2/3} \delta^2 \right] \approx \frac{1}{3}\left(
\frac{U_0}{V}\right) \left[1 + \frac{4}{5} (3\pi^2)^{2/3} \delta^2
\right]. \label{fermiyaqiang-1}
\end{eqnarray}
Thus corrections occur in the equation of state and this could
lead to some cosmological consequences.

\subsection{$\kappa\simeq 1$ and $\kappa \gg 1$ case}

Since the full theory is absent, we take eqn. (\ref{1dGUP}) as the
starting point to gain some features of the quantum theory of
gravity. Therefore, there is effectively no upper limit on $p^2$.
The $\kappa \ll 1$ case corresponds to perturbative region while
$\kappa \ge 1$ is equivalent to $\beta p^2 \ge 1$.

Note that in (\ref{fDefinition}), $f(\kappa)$ is a monotonically
increasing and bounded function of $\kappa$ in the region
$\kappa\geq 0$. So is $N/V$ with respect to $\varepsilon_F$.
At the Hagedorn energy, 
$\kappa=1$ gives $f(\kappa)=\pi/32$. $\kappa\gg 1$ leads to an
upper limit of $f(\kappa)$, which is $\pi/16$. Then at zero
temperature,
\begin{eqnarray}
\frac{N}{V}(\kappa=1)&=&\frac{8\pi }{(hc)^3}E_{\rm H}^3\frac
{\pi}{32}=\frac{1}{32\pi}\frac{1}{\Delta_\textrm{min}^3},
\label{lizishumidujiz1}\\ \frac{N}{V}(\kappa\gg 1)&=&\frac{8\pi
}{(hc)^3}E_{\rm H}^3\frac
{\pi}{16}=\frac{1}{16\pi}\frac{1}{\Delta_\textrm{min}^3}.
\label{lizishumidujiz123}
\end{eqnarray}
From (\ref{fermijitaineng-124}), the energies of the ideal Fermi
gases are
\begin{eqnarray}
U_0(\kappa=1)&=&\frac{8\pi V}{(hc)^3}\frac
{E_{\rm H}^4}{16},\label{nengliangmidujiz1}\\
U_0(\kappa\gg 1)&=&\frac{8\pi V}{(hc)^3}\frac {E_{\rm
H}^4}{4}.\label{nengliangmidujiz1234}
\end{eqnarray}
With (\ref{yaqheneinengguanxi-1121}), the pressures are
\begin{eqnarray}
P_0(\kappa=1)&=&\frac{\pi-2}2\frac{1}V\;U_0(\kappa=1),\label{kdengyu1dpressures}\\
P_0(\kappa\gg
1)&=&\left(\frac{\pi}4\kappa-1\right)\frac{1}V\;U_0(\kappa\gg
1).\label{kdengyu1dpressures22}
\end{eqnarray}
From (\ref{lizishumidujiz123}) and (\ref{nengliangmidujiz1234}),
we can see that the particle number density and the energy density
are bounded from above by the minimal length. However, from
(\ref{kdengyu1dpressures22}), with the increasing of the Fermi
energy, the Fermi degenerate pressure blows up. This divergence
indicates that more and more energy is needed to push a particle
into the system. Furthermore, the equations of state are also
altered.

\section{Ultra-compact stars at zero temperature}
For white dwarfs, the major contribution to the mass is not from
the electron gas but from the non-relativistic cold nuclei:  $M=
2Nm_p$ where $m_p$ is the mass of a proton. There may exist other
configurations that the star is almost composed of
ultra-relativistic particles. In this scenario, the mass is
expressed as $M=U_0/c^2$.

\subsection{White dwarfs with $M=2 N m_p$}
The condition $M= 2Nm_p$ indicates that $\varepsilon_F \ll m_p
c^2$. Therefore, $\kappa \ll 1$ for white dwarfs. The equation of
state (\ref{fermiyaqiang-1}) shows a significant fact that the
quantum gravity effects increase the degenerate pressure. This in
turn induces corrections to the mass of white dwarfs.
A typical model of a white dwarf has the following two properties
\cite{RKPathria1972SM}:
\begin{itemize}
    \item The dynamics of the electrons is relativistic.
    \item the electron gas is completely degenerate and can be
    treated as a zero temperature gas.
\end{itemize}
In $3$-dimensional space, we consider a star with mass $M$ and $N$
electrons. In the absence of gravitation, it is necessary to have
``external walls'' to keep the gas at a given density. For macro
stars such as white dwarfs, it is the electronic degeneracy
pressure to resist the gravitational collapse. That is, on
equilibrium,
\begin{equation}
P_0(R)=\frac{\alpha}{4\pi}\frac{GM^2}{R^4},
\label{pinghengyaqiangyinqi-1}
\end{equation}
where $R^3\sim V$, $G$ is the constant of gravitation and $\alpha$
is a number (of the order of unity) whose exact value depends upon
the distribution of matter inside stars. From eqn.
(\ref{fermiyaqiang-1}) and eqn. (\ref{pinghengyaqiangyinqi-1}), we
have
\begin{equation}
\left(\frac
NV\right)^{4/3}hc\left(1+\delta^2\right)=\frac{GM^2}{R^4}
,\label{pinghe23423angyinqi-1}
\end{equation}
where constants of the order of unity are ignored. Plugging in $M=
2N m_p$, the correction to the mass of stars is given by
\begin{equation}
M=M_0\left[1+\delta^2\right]=M_0\left[1+\left(\frac
NV\right)^{2/3}\beta h^2\right],\;\;\;M_0=\left(\frac
1{2m_p}\right)^2\left(\frac{hc}{G}\right)^{\frac 32}\approx
M_{\odot}. \label{W-D-C2222er}
\end{equation}
Here, $M_0$ is the mass of a white dwarf without
corrections\footnote{The relation between the mass and the radius
of white dwarfs is calculated with the dispersion $\varepsilon = m
c^2 \Big[ \sqrt {1+ \left(\frac{p}{mc}\right)^2} -1\Big]$. Since
we are only concerned with the mass limit of the star
configuration, the ultra-relativistic dispersion $\varepsilon =
cp$ is enough. This is shown by $M_0$ in eqn.
(\ref{W-D-C2222er}).}.
For white dwarfs, $M_0$ is approximately equal to the
Chandrasekhar limit, which is $1.44$ times the mass of the sun
$M_{\odot}$. One can see that the quantum gravity correction
depends on the number density of the star. For a typical white
dwarf, the number density of electrons is
$10^{36}\;\textrm{m}^{-3}$, which gives the average distance
$\bar{d}=10^{-12}\textrm{m}$. The Fermi energy is about
$10^5\;\textrm{eV}$ (far exceeds the white dwarf's central
temperature $T$ $\sim 10^7\,\textrm{K}$). Then
(\ref{feimijinsitia-1}) is well satisfied even for the upper limit
of $\beta_0\sim 10^{36}$.
Taking $\beta_0= 10^{36}$, one finds
\begin{equation}
M=M_0\left[1+10^{-10}\right].
\end{equation}
Although the correction is small, one should note that the
existence of quantum gravity correction is positive. Therefore,
the quantum effect of gravity  tends to resist the collapse of
stars. Furthermore, larger $\beta_0$ leads to more obvious effects
of quantum gravity.

\subsection{Configurations with $M = U_0/c^2$}

We consider a compact star entirely constituted by an  ideal Fermi
gas.
The total mass is given by eqn. (\ref{fermijitaineng-124}),
$M=U_0/{c^2}$. The equilibrium equation
(\ref{pinghengyaqiangyinqi-1}) leads to
\begin{equation}
R=\frac {\alpha}{24}R_S^0\;\tilde{h}(\kappa), \quad\quad \tilde h
(\kappa)\equiv\frac{\kappa^4}{(1+\kappa^2)^2 g(\kappa)}\,,
\label{zhongweisdfsdfdso1-9}
\end{equation}
where $R_S^0$ is the Schwarzschild radius of the star with mass
$M=U_0/{c^2}$.
\begin{equation}
R_S^0=\frac{2GM}{c^2}= \frac{2GU_0}{c^4}.
\end{equation}

\begin{figure}
\centerline{\hbox{\epsfig{figure=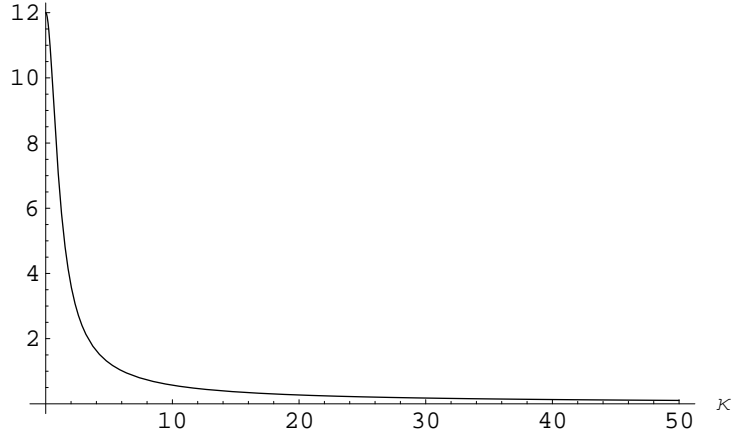, height=6cm}}}
\caption{The $\tilde h(\kappa)$ in eqn.
(\ref{zhongweisdfsdfdso1-9})  versus $\kappa$. $\tilde h(\kappa)$
is a bounded and approaches zero at infinity.} \label{FigTildeH}
\end{figure}

\noindent As shown in Figure \ref{FigTildeH}, $\tilde{h}(\kappa)$
is a monotonically decreasing function of $\kappa$ with $\tilde
h(0) =12$ and $\tilde h (\infty) =0$. Therefore, at a small value
of $\kappa$, formation of a compact star is triggered. While at a
critical value of $\kappa=\kappa^*$ which makes
$R(\kappa^*)=R^0_S$, a black hole is created and thereafter. To
determine $\kappa^*$, one has to know $\alpha$. It is conceivable
that better precision can be achieved when  general relativity is
taken into account in the equilibrium equation
(\ref{pinghengyaqiangyinqi-1}). What is interesting is that inside
the horizon, it is not a singularity but an ensemble of highly
degenerate fermion gas.

\begin{figure}
\centerline{\hbox{\epsfig{figure=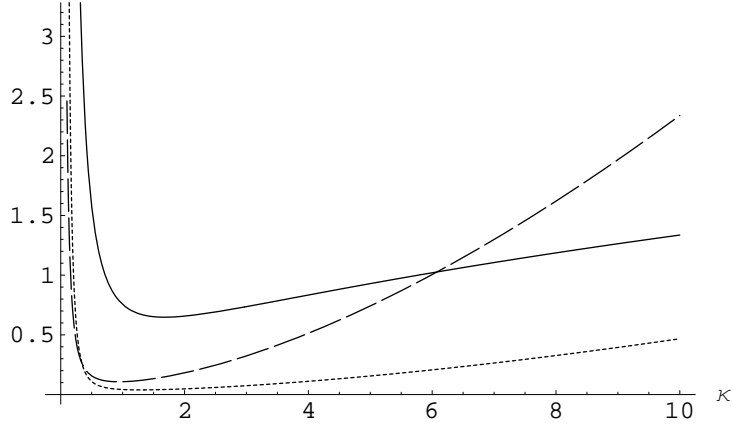, height=6cm}}}
\caption{For a star configuration, the radius $R$ (solid line) in
eqn. (\ref{zhongweizibieo1-9}), the total number of particles $N$
(dotted line) in eqn. (\ref{totalNabcd}) and the mass $M$ (dashed
line) in eqn. (\ref{masscurveaa}) versus $\kappa$. All of them are
divergent at $\kappa=0$ and $\kappa\to\infty$. The radius has a
minimum around $\kappa=1.681$. The total number of particles
reaches a minimum around $\kappa=1.290$. The mass acquires a
minimum around $\kappa=0.940 $.} \label{FigHNM}
\end{figure}

It is useful to write eqn. (\ref{zhongweisdfsdfdso1-9}) in the
form:
\begin{equation}
R^2=\frac {36\pi}\alpha\beta_0^2 \ell_p^2\; h(\kappa), \quad\quad
h(\kappa)\equiv g(\kappa)
\frac{(1+\kappa^2)^4}{\kappa^8}.\label{zhongweizibieo1-9}
\end{equation}
Refer to the solid line in Figure \ref{FigHNM}, the radius is
divergent at $\kappa=0$ and $\kappa\to\infty$. There is a minimum
$h_{\rm min}=0.420$ around $\kappa=1.681$.
Thus a minimum radius is given by
\begin{equation}
R_{\textrm{min}}\simeq \beta_0
\ell_p=\sqrt{\beta_0}\Delta_{\textrm{min}},\label{kabawei1djixianbanj}
\end{equation}
which implies that no singularity is present. This conclusion is a
consequence of the existence of the minimum length. Contradicting
to the intuition, the minimum radius is not proportional to the
minimum length but to $\sqrt{\beta_0}\Delta_{\textrm{min}}$.
Nevertheless, $\beta_0$ may be much larger than unity as proposed
by many models. Whether a star of the minimum radius is a black
hole depends on the value of $\alpha$ since $\tilde h(1.681)\simeq
4.365$.

Substituting (\ref{zhongweizibieo1-9}) into
(\ref{fermilizishu-1}), we obtain the total number of particles,
the dotted line in Figure \ref{FigHNM},
\begin{equation}
N=288\sqrt\pi\left(\frac{1}{\alpha}
\right)^{3/2}\beta_0^{3/2}f(\kappa)h^{3/2}(\kappa).
\label{totalNabcd}
\end{equation}
In the vicinity of $\kappa=1.290$, $N$ reaches a minimal value
\begin{equation}
N_{\textrm{min}}=N(\kappa=1.290)=0.0392\times
288\sqrt\pi\left(\frac{1}{\alpha}\right)^{3/2}\beta_0^{3/2}\sim
\beta_0^{3/2}.
\end{equation}
Substituting (\ref{zhongweizibieo1-9}) into
(\ref{fermijitaineng-124}), one acquires the total mass, the
dashed line in Figure \ref{FigHNM},
\begin{equation}
M=\frac{U_0}{c^2}= \frac {72\sqrt\pi}{\alpha^{3/2}} \beta_0 M_p\;
h(\kappa)^{3/2}\frac{\kappa^4}{(1+\kappa^2)^2}.
\label{masscurveaa}
\end{equation}
Around $\kappa=0.940 $, $M$ has a minimum
$$
M_{\textrm{min}}=M(\kappa=0.940)=0.107\times \frac 2{\sqrt{\alpha
\pi}} M_p\, \beta_0 \approx 10^{-8}\beta_0 \;(\textrm{kg}).
$$

The divergencies of the radius, total number of particles and mass
at $\kappa=0$ are not caused by quantum gravity effects.
Expanding the radius, total number of particles and mass for small
$\kappa$ respectively:
\begin{equation}
R = \sqrt\frac{3\pi}{\alpha} \frac{(c\hbar)^2}{\ell_p}
\frac{1}{\varepsilon_F^2} \left[ 1 + \beta_0 \frac{7}{5} \left(
\frac{\varepsilon_F}{M_p c^2}\right)^2 + O(\kappa)^4\right].
\end{equation}
\begin{equation}
N= \frac{1}{4 \alpha^{3/2}} \sqrt\frac{\pi}{3}
\left(\frac{c\hbar}{\ell_p}\right)^3 \frac{1}{\varepsilon_F^3}
\left[ 1 + \beta_0 \frac{12}{5} \left( \frac{\varepsilon_F}{M_p
c^2}\right)^2 + O(\kappa)^4\right].
\end{equation}
\begin{equation}
M=\frac{\sqrt{3\pi}}{\alpha^{3/2}}M_p^3
c^4\frac{1}{\varepsilon_F^2} \left[ 1 + \beta_0 \frac{11}{5}
\left( \frac{\varepsilon_F}{M_p c^2}\right)^2 +
O(\kappa)^4\right].
\end{equation}
The quantum gravity factor $\beta_0$ has no presence on the
leading orders and only shows as small corrections. In fact, the
ultra-relativistic dispersion $\varepsilon = cp$ is only
applicable to the situation $\varepsilon_F\gg mc^2$, where $m$ is
the rest mass of the particles building the star. Therefore, our
results are valid in the region $\kappa\gg \sqrt \beta_0
\frac{m}{M_p}$.
When investigating the region $\varepsilon_F \leq mc^2$, the
relativistic dispersion or the non-relativistic one ought to be
used and one also has $\kappa\ll 1$.

As $\kappa$ keeps increasing, the Fermi pressure monotonically
blows up to resist the gravitational collapse. Though both the
total number of particles and mass are divergent at large
$\kappa$, the number density and mass density approach finite
quantities as discussed in last section. From eqn.
(\ref{zhongweisdfsdfdso1-9}) and Figure \ref{FigTildeH}, beyond
the critical $\kappa^*$, the star behaves as a black hole, in a
rough definition. Eventually, the radius approaches infinity, and
yet its ratio to the Schwarzschild one vanishes.

It is of interest to have a look at the asymptotic  behavior of
the pressure in terms of the radius. For large $\kappa$,
\[
P_0 = \frac{\pi^2}{2}\frac{E_{\rm H}^4}{(hc)^3}\, \kappa +
O(1),\quad R= \frac{3\pi}{2\sqrt\alpha} \beta_0\ell_p \sqrt\kappa
+ O\left(\frac{1}{\sqrt\kappa}\right).
\]
Then,
\begin{equation}
\frac{dP_0}{dR}=\frac{\alpha}{9\pi^2}\left(\frac{c}{h}\right)
\frac{1}{\beta_0^4\ell_p^6}\, R \quad \quad {\rm as}\quad
\kappa\gg 1. \label{RressureAndRadiusL}
\end{equation}
Therefore, the increment of the Fermi pressure is accelerated for
larger radius. If take the viewpoint that the whole universe is a
big black hole, supported by low temperature permeating Fermi
gases, the resulted Fermi pressure may provide a mechanism of the
accelerated expansion. We anticipate that future works which make
use of general relativity and interacting systems can predict
observationally consistent results.

It is of importance to study the stability of the configurations.
The strategy is to fix the total number of particles $N$ and vary
the radius. To have a stable configuration, a positive (negative)
perturbation of the radius must make the degenerate pressure
smaller (larger) than the gravitational pressure.
From eqns
(\ref{fermilizishu-1}), (\ref{fermijitaineng-124}) and
(\ref{yaqheneinengguanxi-1121}), we have
\begin{equation}
P_0 - \frac{\alpha}{4\pi}\frac{G M^2}{R^4} =  const\,
\frac{\kappa^8}{(1+\kappa^2 )^4} f^{-2/3}(\kappa) \big( j(\kappa)
- j(\kappa_E) \big),\quad j(\kappa) \equiv h(\kappa)
f^{2/3}(\kappa), \label{StableJudge}
\end{equation}
where $\kappa_E$ is the  value of $\kappa$ on the equilibrium for
a fixed $N$. From eqn. (\ref{fermilizishu-1}), a small
perturbation of the radius leads to a shift of $\kappa$ with an
opposite sign:
\[
R \to R \pm |\Delta R| \Rightarrow \kappa \to \kappa \mp |\Delta
\kappa|.
\]
With eqn. (\ref{StableJudge}), one readily finds that in order to
have stable configurations, $j(\kappa)$ must be an increasing
function of $\kappa$.  It is easy to see that $j(\kappa)$ has the
same behavior as the dotted line, total number of particles $N$,
in Fig \ref{FigHNM}. Therefore, there exists another critical
value $\kappa_S \cong 1.290$. The star configurations are stable
as $\kappa > \kappa_S$. The minimum radius is obviously located in
the stable region. The instability of the configurations as $0
\ll\kappa \leq \kappa_S$ may not be very true. To have more
reliable results, one should make use of the relativistic
dispersion.

It is worthwhile to take a glance at a compact star comprised
primarily of fermions with a small portion of bosons. This
configuration is different from the two structures we have
explored in this section. There does not exist any constraint
between the mass and fermion number or energy. From the
equilibrium condition eqn. (\ref{pinghengyaqiangyinqi-1}), as more
bosons are inhaled, the degenerate pressure $P_0$ and $\kappa$
move up. With eqn. (\ref{fermilizishu-1}), the volume shrinks to
its minimum
\begin{equation}
V_{\textrm{min}}=16\pi N\Delta_{\textrm{min}}^3, \label{FBMV}
\end{equation}
which is obviously a consequence of the existence of the minimum
length. One can easily conceive that a black hole, without
singularity, is going to form as the portion of bosons is growing.

\section{Summary and Discussion}

Based on the minimum observable length, we discussed the quantum
gravity influences on the statistic properties of ideal gases. The
finite temperature classical non-relativistic, ultra-relativistic
and photon gases were addressed. Small corrections to the energy
and entropy at low temperatures were found. The temperature itself
is unaltered. Moreover, modifications to the equations of state
may have some cosmological consequences. When connected to the
traditional grand-unified string model, an lower bound
$\beta_0>10^4$ was given.

We then paid attention to the ultra-relativistic Fermi gases at
zero temperature. The energy density and pressure receive positive
corrections proportional to $\hbar^2\beta (N/V)^{2/3}$ in the case
of $\kappa\ll 1$. When applied to the white dwarfs, the
corrections tend to resist the gravitational collapse of stars and
then lift up the Chandrasekhar limit. Taking the GUP as a starting
point, we also studied the high Fermi energy situation. It is
shown that both the number density and the energy density approach
finite values, whereas the degenerate pressure blows up for large
$\kappa$.

For compact stars completely constituted by an ultra-relativistic
ideal Fermi gas, we found that the radius, total number and mass
all achieve their minima around $\kappa\simeq 1$. The minimum
radius is the order of $\sqrt\beta_0 \Delta_{\rm min}$. As
$\kappa$ increases from a small value, a compact star is formed.
After passing through a critical value $\kappa=\kappa^*$, where
the radius coincides its Schwarzschild one, the compact star
behaves as a black hole. Eventually, the radius approaches
infinity, nevertheless the ratio to its Schwarzschild radius goes
to zero. In this sense, the universe may be interpreted as a giant
black hole. Furthermore, at large $\kappa$, the degenerate
pressure increases at a rate proportional to the radius. This
result provides a possible account for the accelerated expansion
of the universe. The configurations are stable beyond another
critical Fermi energy $\kappa_S \cong 1.290$.

There are several questions that we have not addressed. The first
one is that in calculating the pressure balance of compact stars,
we employed the Newton's gravity. At large $\kappa$,  as we
emphasized, it is more reasonable to include the corrections from
general relativity. For instance, replace the equilibrium
condition  eqn. (\ref{pinghengyaqiangyinqi-1}) by the
Tolman-Oppenheimer-Volkoff (TOV) equation. Moreover, as $\kappa$
is very large, ideal gas model only serves as a leading order
approximation. Better refined statistic models can certainly offer
more accurate predictions. In the studies of Fermi gases and
compact star configurations, we adopted zero temperature and the
ultra-relativistic dispersion.
It is of interest to figure out a relation between the
mass and radius with relativistic dispersions . Furthermore, more
precise results are expected by assuming low but nonzero
temperatures, though the calculations would be much more
complicated.

We chose eqn. (\ref{statenonumber-LNC1}) to realize quantum
gravity effects. However, whether Liouville's theorem is
applicable in the high temperature limit or ultra-high density
lacks of proof in principle. Atick and Witten have shown that at
temperatures far above the Hagedorn temperature, string theory has
a very peculiar thermodynamic behavior
\cite{AtickWitten1988NPB310}. On the other hand, the space-time
dimensionality may be reduced  at short distances
\cite{Ambjorn2005PRL}. Therefore we naturally expect that the
dimension $d$ in the equation (\ref{statenonumber-LNC1}) may run
with energy scales. The establishment of a specific model to show
the phase transition at the Hagedorn temperature will be discussed
in follow-up works.


\section*{Acknowledgement}

We thank Dr. X. Guo, J. Tao,  B. Mu and H. Wu for the useful
discussions. This work is partially supported by the Fundamental
Research Funds for the Central Universities (Grant No.
ZYGX2009X008),  NSFC (Grant No.10705008) and NCET.



\begin{thebibliography}{1}

\small

\bibitem{Veneziano1986EPL199}
G. Veneziano, \emph{A stringy nature needs just two constants},
Europhys. Lett. \textbf{2} (1986) 199.
%
\bibitem{DJGross1988NPB407}
D. J. Gross and P. F. Mende, \emph{String theory beyond the Planck
scale}, Nucl. Phys. \textbf{B 303} (1988) 407.
%
\bibitem{DAmati1989PLB41}
D. Amati, M. Ciafaloni and G. Veneziano, \emph{Can space-time be
probed below the string size?} Phys. Lett. \textbf{B 216} (1989) 41.
%
\bibitem{KKonishi1990PLB276}
K. Konishi, G. Paffuti and P. Provero, \emph{Minimum physical length
and the generalized uncertainty principle in string theory}, Phys.
Lett. \textbf{B 234} (1990) 276.
%
\bibitem{RGuida1991MPL1487}
R. Guida, K. Konishi and P. Provero, \emph{On the short distance
behavior of string theories}, Mod. Phys. Lett. \textbf{A 6} (1991)
1487.
%
\bibitem{Maggiore1993PLB304}
M. Maggiore, \emph{A generalized uncertainty principle in quantum
gravity}, Phys. Lett. \textbf{B 304} (1993) 65 [hep-th/9301067].
%
\bibitem{Maggiore1993PLB319}
M. Maggiore,  \emph{The algebraic structure of the generalized
uncertainty principle}, Phys. Lett. \textbf{B 319} (1993) 83
[hep-th/9309034].
%
\bibitem{Garay1995IJMPA145}
L. J. Garay, \emph{Quantum gravity and minimum length}, Int. J. Mod.
Phys. \textbf{A 10} (1995) 145 [gr-qc/9403008].
%
\bibitem{Kempf1995PRD1108}
A. Kempf, G. Mangano and R. B. Mann, \emph{Hilbert space
representation of the minimal length uncertainty relation}, Phys.
Rev. \textbf{D} \textbf{52} (1995) 1108 [hep-th/9412167].
%
\bibitem{Kempf1992LMP26}
A. Kempf, \emph{Quantum group-symmetric fock spaces with
bargmann-fock representation}, Lett. Math. Phys. \textbf{26} (1992)
1.
%
\bibitem{Kempf199234JMP969}
A. Kempf, \emph{Quantum group Symmetric Bargmann-Fock space:
integral kernels, Green functions, driving forces}, J. Math. Phys.
\textbf{34} (1994) 969.
%
\bibitem{Kempf199235JMP4483}
A. Kempf, \emph{Uncertainty relation in quantum mechanics with
quantum group symmetry}, J. Math. Phys. \textbf{35} (1994) 4483
[hep-th/9311147].
%
\bibitem{Hossenfelder2006CQG1815}
S. Hossenfelder, \emph{A note on theories with a minimal length},
Class. Quantum Grav. \textbf{23} (2006) 1815.
%
\bibitem{Hossenfelder2008CQG038003}
S. Hossenfelder, \emph{A note on quantum field theories with a
minimal length scale}, Class. Quantum Grav. \textbf{25} (2008)
038003.
%
\bibitem{RAMA2001PLB}
S. K. Rama, \emph{Some consequences of the Generalised Uncertainty
Principle: Statistical Mechanical, Cosmological and Varying Speed of
Light}, Phys. Lett. \textbf{B 519} (2001) 103 [hep-th/0107255].
%
%
\bibitem{Fityo2008PLA}
T. V. Fityo, \emph{Statistical physics in deformed spaces with
minimal length}, Phys. Lett. \textbf{A 372} (2008) 5872
[quan-th/07120891].
%
\bibitem{DasVagenas2008PRL}
S. Das and E. C. Vagenas, \emph{Universality of quantum gravity
corrections}, Phys. Rev. Lett. \textbf{101} (2008) 221301.
%
\bibitem{FBrauB2006PRD036002}
F. Brau and F. Buisseret, \emph{Minimal length uncertainty relation
and gravitational quantum well}, Phys. Rev. \textbf{D 74} (2006)
036002.
%
\bibitem{LNC2002PRD}
L. N. Chang, D. Minic, N. Okamura and T. Takeuchi, \emph{Effect of
the minimal length uncertainty relation on the density of states and
the cosmological constant problem}, Phys. Rev. \textbf{D 65} (2002)
125028 [hep-th/0201017].
%
\bibitem{MuWuYang2009print}
Benrong Mu, Houwen Wu and Haitang Yang, \emph{The generalized
uncertainty principle in the presence of extra dimensions},
arXiv:0909.3635.
%
\bibitem{RAMA2001-0204215}
S. K. Rama, \emph{Dynamical features of maggiore's generalised
commutation relations}, hep-th/0204215.
%
%
\bibitem{NozariMehdipour2007Chaos}
K. Nozari, S. H. Mehdipour, \emph{Implications of minimal length
scale on the statistical mechanics of ideal gas}, Chaos, Solitons
and Fractals, \textbf{32} (2007)1637-1644 [hep-th/0601096].
%
\bibitem{AtickWitten1988NPB310}
J. J. Atick and E. Witten, \emph{The Hagedorn transition and the
number of degrees of freedom of string theory}, Nucl. Phys.
\textbf{B 310} (1988) 291.
%
\bibitem{BZwiebach2004firstcoursestring16}
B. Zwiebach, \emph{A first course in string theory}, Ref. Chap. 16,
Cambridge University Press (2004).
%
\bibitem{RKPathria1972SM}
R. K. Pathria, \emph{Statistical Mechanics}, Pergamon Press, First
Edition, 1972.
%
\bibitem{Ambjorn2005PRL}
J. Ambj{\o}rn, J. Jurkiewicz and R. Loll, \emph{The Spectral
Dimension of the Universe is Scale Dependent}, Phys. Rev. Lett.
\textbf{95} (2005) 171301.
%


\end{thebibliography}
\end{document}